\documentclass{aastex}
\input{epsf}
\usepackage{emulateapj5,apjfonts}

\def\feka{Fe K$\alpha$}
\def\chandra{{\it Chandra}} 
 
\def\asca{{\it ASCA}} 
 
\def\rxte{{\it RXTE}} 
\def\sax{{\it BeppoSAX}}

\def\lum{erg s$^{-1}$}
\def\flux{erg cm$^{-2}$ s$^{-1}$}
\def\nh{cm$^{-2}$}
\def\arcsec{$^{\prime\prime}$}

\def\cir{Circinus}

\def\ltsima{$\; \buildrel < \over \sim \;$}
\def\simlt{\lower.5ex\hbox{\ltsima}} 
\def\gtsima{$\; \buildrel > \over \sim \;$}
\def\simgt{\lower.5ex\hbox{\gtsima}} 

\slugcomment{Accepted for publication in the Astrophysical Journal Letters}

\shorttitle{CHANDRA HETGS SPECTROSCOPY OF CIRCINUS}

\shortauthors{SAMBRUNA ET AL.}

\begin{document}

\title{High-resolution X-ray spectroscopy of the Seyfert 2 galaxy
Circinus with \chandra}

\author{Rita M. Sambruna,\altaffilmark{1} 
Hagai Netzer,\altaffilmark{2} 
Shai Kaspi,\altaffilmark{1}
W. N. Brandt,\altaffilmark{1}
G.Chartas,\altaffilmark{1}
G.P.Garmire,\altaffilmark{1}
John A. Nousek,\altaffilmark{1}
and
K.A.Weaver,\altaffilmark{3}
}
\altaffiltext{1}{Department of Astronomy and Astrophysics, 525 Davey
Laboratory, The Pennsylvania State University, University Park, PA 16802.}
\altaffiltext{2}{School of Physics and Astronomy, Raymond and Beverly
Sackler Faculty of Exact Sciences, Tel-Aviv University, Tel-Aviv 69978, Israel.}
\altaffiltext{3}{Laboratory for High Energy Astrophysics, Code 660,
NASA/Goddard Space Flight Center, Greenbelt, MD 20771.}

\begin{abstract} 

Results from a 60 ks \chandra\ HETGS observation of the nearby Seyfert
2 \cir\ are presented. The spectrum shows a wealth of emission lines
at both soft and hard X-rays, including lines of Ne, Mg, Si, S, Ar,
Ca, and Fe, and a prominent \feka\ line at 6.4 keV. We identify
several of the He-like components and measure several of the Lyman
lines of the H-like ions.  The lines' profiles are unresolved at the
limited signal-to-noise ratio of the data. Our analysis of the
zeroth-order image in a companion paper constrains the size of the
emission region to be 20--60 pc, suggesting that emission within this
volume is almost entirely due to the reprocessing of the obscured central
source. Here we show that a model containing two distinct components
can reproduce almost all the observed properties of this gas. The
ionized component can explain the observed intensities of the ionized
species, assuming twice-solar composition and an $N \propto r^{-1.5}$
density distribution.  The neutral component is highly concentrated,
well within the 0.8\arcsec\ point source, and is responsible for
almost all of the observed K$\alpha$ (6.4 keV) emission. \cir\ seems
to be different than Mkn~3 in terms of its gas distribution.

\end{abstract}

\keywords{
galaxies: active --- 
galaxies: nuclei --- 
galaxies: Seyfert --- 
galaxies: individual (Circinus) --- 
X-rays: galaxies}

\section{Introduction}

Recent X-ray studies of Seyfert 2 galaxies with \asca\ and \sax\ have
shown that the 0.1--10 keV spectra of these sources are rich in
emission lines at both soft and hard energies (e.g., Guainazzi et
al. 1999; Turner et al. 1997). The interpretation of the emission
lines is problematic because of ambiguities about line blending,
line profiles, and line flux distribution that are all poorly
constrained by \asca\ and \sax. The data are consistent with emission
from gas in photoionization equilibrium (e.g., Netzer, Turner, \&
George 1998 and references therein), but there are also attempts to
fit the spectra by a two-temperature gas in collisional equilibrium
(Ueno et al. 1994), presumably due to starburst emission. X-ray
observations at high resolution both spatially and spectrally are
crucial to determining the origin of the X-ray lines in Seyfert 2s, a
task for which \chandra\ is uniquely suited.

Here we present a 60 ks \chandra\ HETGS spectrum of the Seyfert 2
galaxy \cir. Previous \asca\ and \sax\ observations of this source
revealed several emission lines at both soft and hard energies (Sako
et al. 2000a; Guainazzi et al. 1999; Matt et al. 1996). In a companion
paper, focusing on the zeroth-order ACIS image, we established that
several components contribute to the X-ray emission from \cir. In
particular, we found that $\sim$ 60\% of the X-ray flux at \simlt 2
keV is due to an extended component on scales $\sim$ 2.3\arcsec, while
at harder energies the contribution from a compact (\simlt 0.8\arcsec)
region prevails. Importantly, the ACIS spectrum of the latter
component exhibits several emission lines including a prominent (EW
$\sim$ 2.5 keV) \feka\ line. In this paper, we concentrate on the
analysis of the HETGS spectrum and on the implications for the
physical conditions of the emitting gas. We also use a simultaneous
\rxte\ observation to derive useful constraints on the intrinsic
nuclear X-ray continuum. At the distance of the galaxy ($\sim$ 4 Mpc),
1\arcsec=19 pc.

\section{Observations and Data Analysis} 

\cir\ was observed with the High Energy Transmission Grating
Spectrometer (HETGS; Canizares et al. 2000, in prep.) on 2000 June 6,
with ACIS-S (Garmire et al. 2000) in the focal plane. The total net
exposure was 60,223 s. Details on the observation are given in our
companion paper.
 
The HETGS carries two mirror assemblies, the High Energy Grating (HEG)
and Medium Energy Grating (MEG). The nuclear HEG and MEG spectra were
extracted in a narrow (15 pixel) rectangular region centered on the
zeroth-order position, avoiding contamination from the dispersed
spectra of the nearby serendipitous sources.  Although a few of the
serendipitous sources exhibit emission lines in their ACIS spectra at
both soft and hard X-rays, the 0.5--8 keV flux of the brightest one is
a factor 5 weaker than the nucleus. We thus believe that contamination
to the HETGS nuclear spectrum (in the regions of overlap of the
dispersed spectra) is negligible. The HEG and MEG spectra were gain
corrected and flux calibrated using Ancillary Response Files generated
with the \verb+CIAO+ software. Only the first order HEG and MEG
spectra were used for the analysis, as higher orders contain only a
few (\simlt 3 per bin) counts and are not useful.  The spectra were
also corrected for cosmological redshift and Galactic absorption,
$N_{\rm H}^{Gal}=3.3 \times 10^{21}$ \nh\ (Freeman et al. 1977). 

\section{Simultaneous \rxte\ observations} 

We used a simultaneous 30 ks exposure with \rxte\ to constrain the
higher-energy X-ray continuum emission from \cir. The \rxte\ data were
reduced following standard criteria; here we use only data from the
PCA detector and report only on the results most relevant for the
modeling of the \chandra\ HETGS data, leaving more details to a
future paper.  The source

\centerline{\epsfxsize=10.5cm\epsfbox{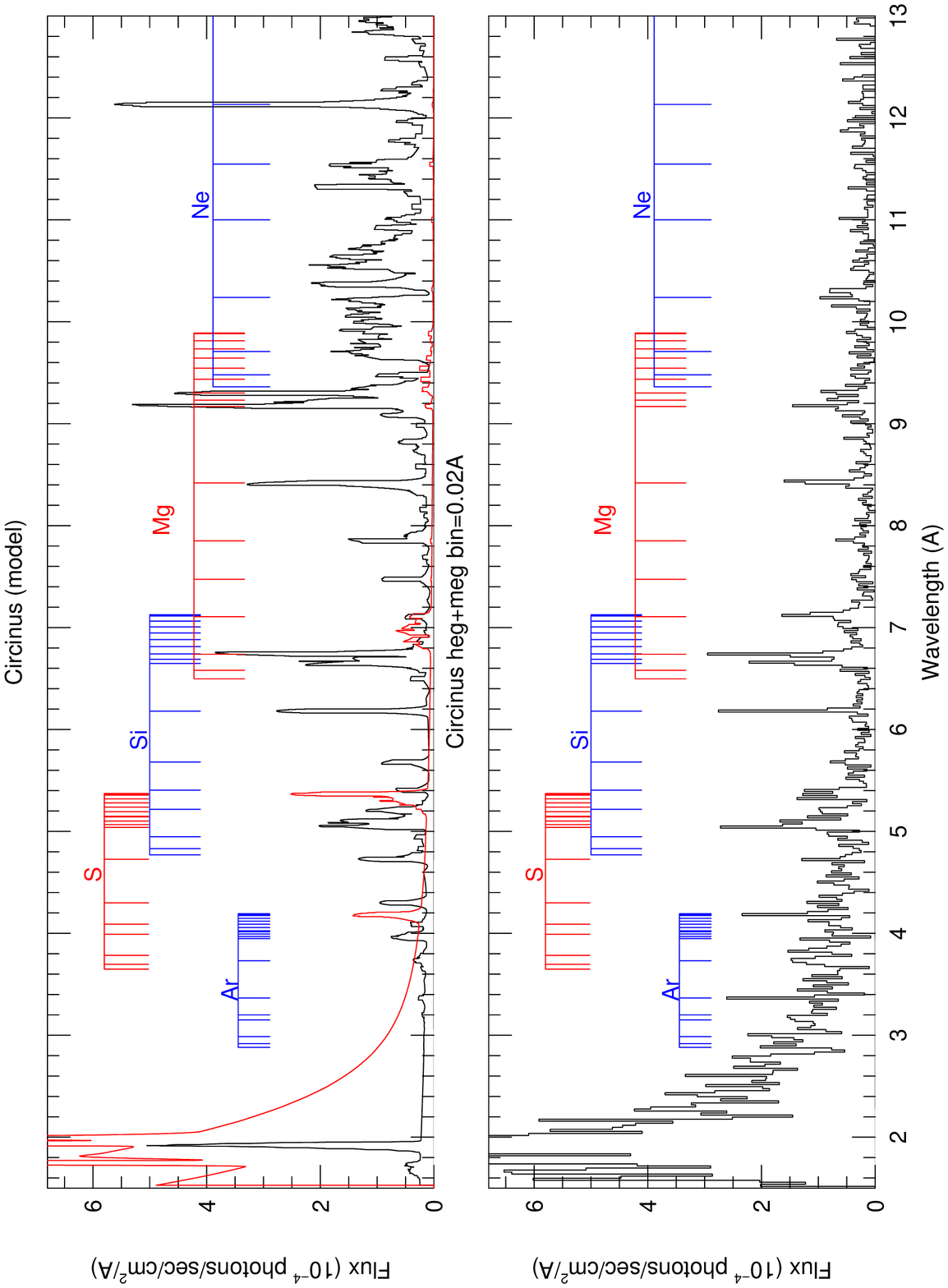}}
\vglue-0.8cm
\figcaption{ Bottom: The X-ray spectrum of \cir\ from a 60 ks \chandra\
HETGS observation. The spectrum was obtained by averaging the MEG and
HEG first-order spectra (after rebinning the HEG data to the same
resolution as the MEG data), and rebinning to 0.02 \AA. Many emission
lines mostly due to H- and He-like elements are detected, together
with weaker lines from ``neutral'' S, Si, and Ar. The most prominent
feature is the \feka\ line at 6.4 keV.  Top: A two component model fit
to the data; ionized component (solid line) and neutral component
(dashed line).
\label{model-data} }
\centerline{}
\centerline{}

\noindent was detected with the PCA up to $\sim$ 30
keV, with a 2--30 keV count rate of 7.26 $\pm$ 0.04 counts s$^{-1}$. 

We fitted the PCA data in the energy range 6--30 keV, where relatively
few lines are present (mainly \feka\ at 6.4 keV, FeK$\beta$ at 7.1
keV, and \ion{Fe}{26}/NiK$\alpha$ at 7.9 keV; Guainazzi et al. 1999).
We find that an excellent description of the PCA data ($\chi^2$=56 for
62 degrees of freedom) is obtained with a model including a ``pure''
reflection continuum from neutral gas, plus a heavily absorbed
($N_{\rm H} \sim 6 \times 10^{24}$ \nh) power law dominant at \simgt
10 keV, plus the three Gaussians lines. The best-fit model and fitted
parameters are in complete agreement with a previous \sax\ observation
of \cir\ (Matt et al. 1999, Guainazzi et al. 1999). We mainly stress
here the results for the intrinsic nuclear continuum: power-law photon
index $\Gamma=1.65^{+1.25}_{-0.35}$ (90\% confidence errors) and
intrinsic (absorption-corrected) 2--10 keV luminosity in the range
$L_{2-10~{\rm keV}}^{intr}=3 \times 10^{40}-3\times 10^{42}$ \lum, in
agreement with the \sax\ data (Matt et al. 1999). 


\section{The HETGS Spectrum} 

The first order MEG spectra agree well with each other within the
resolution of the grating ($\sim$ 0.023 \AA, twice as for the
HEG). The two HEG spectra also agree with each other and with the MEG
spectra. Therefore, in order to increase the signal-to-noise ratio,
the four spectra were averaged after rebinning the HEG data to the MEG
resolution. While this procedure sacrifices the higher resolution of
the HEG, in most cases the lines are unresolved and the loss of
resolution is thus well compensated by a cleaner detection of the
lines.

Figure \ref{model-data} shows a wide wavelength coverage view of the
flux-calibrated spectrum, binned at 0.02 \AA.  A plethora of emission
lines are apparent in the spectrum at all energies. The most prominent
one is the \feka\ line at 6.4 keV, with an EW $\sim 2.1$ keV, an
unresolved core, and a hint of a broader-base component which will be
discussed elsewhere. A detailed view of the   

\centerline{\epsfxsize=10.5cm\epsfbox{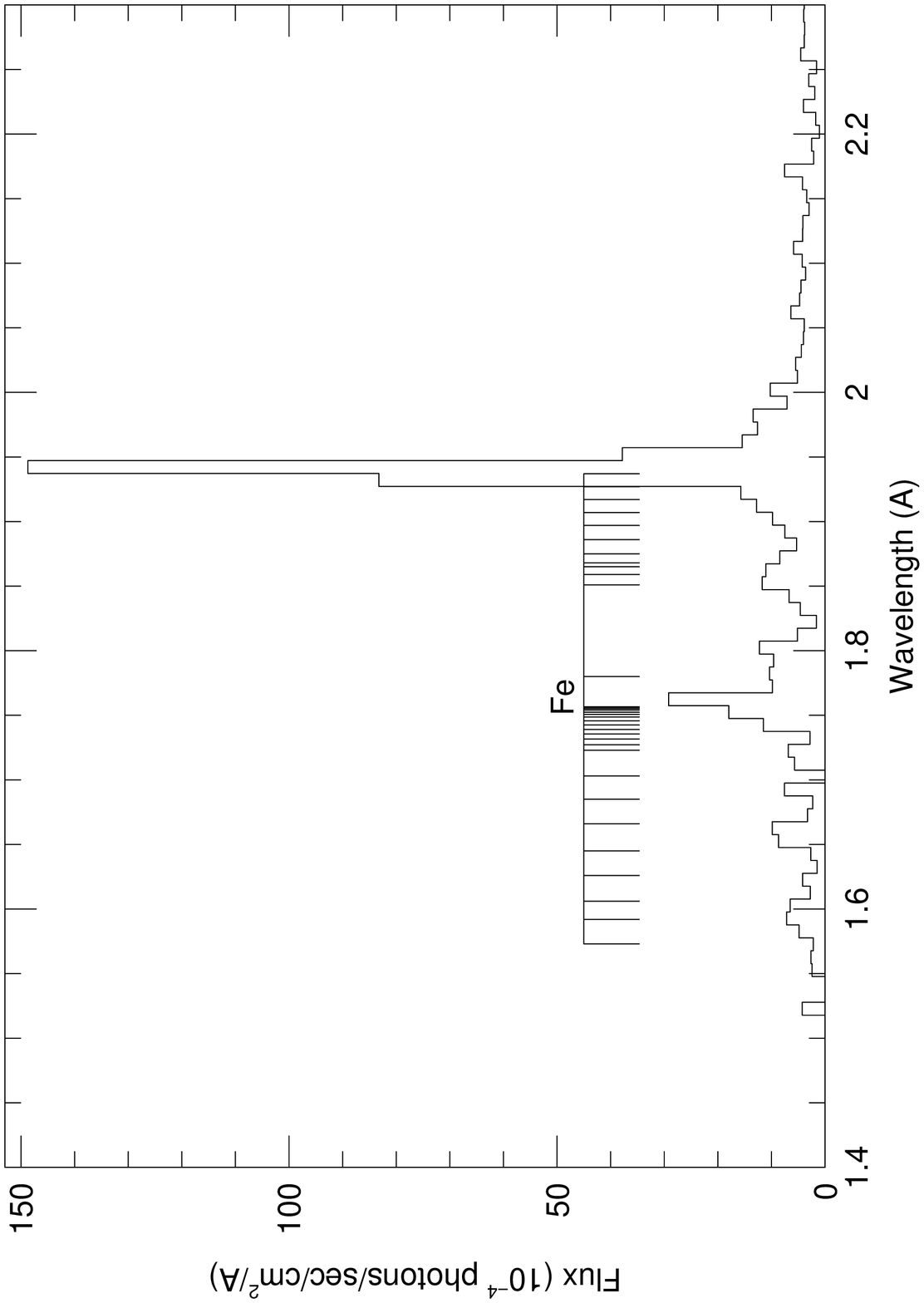}}
\vglue-0.8cm
\figcaption{
The Fe line complex in \cir\ from a 60 ks HETGS exposure shown, this time, 
after rebinning to 0.01 \AA.
The \feka\ line consists of a narrow, unresolved core plus a broader base. Also
detected is the FeK$\beta$ line at 7.1 keV, and a weak line at
6.6 keV, with an energy consistent with \ion{Fe}{25}. 
\label{feline} }
\centerline{}
\centerline{}

\noindent Fe line complex (Figure
\ref{feline}) shows also the Fe K$\beta$\,line at $\sim$ 7.1 keV, with
an intensity of $\sim 0.2$ that of the \feka\ line. We have identified
another feature at around 6.6 keV, consistent with the \ion{Fe}{25}
K$\alpha$ line.  As shown below, our model cannot explain the
intensity of this feature.  The high absorption column in the
direction to \cir, and the relatively faint X-ray flux of the source,
do not allow any clear line detections below $\sim 1$ keV.

Table 1 gives quantitative information about the detected emission
lines, including the line identification (note some question marks due
to uncertain identifications), measured fluxes, and EWs.  Many H-like
and He-like lines of Mg, Si, S and possibly Ar are detected in the
spectrum, together with weaker ``neutral'' lines of Si, S, Ar, and
perhaps Ca.  The high resolution allows deblending of the forbidden,
intercombination, and resonance lines of several of the He-like ions
(Figure \ref{model-data}).  The significance of the lines' detections
can be judged from the corresponding uncertainties on the line fluxes.
The poor signal-to-noise ratio did not allow a meaningful line-width
determination, and all FWHMs are consistent with the instrumental
resolution.

The EWs are calculated with respect to the total observed continuum,
which was evaluated using two independent methods.  First, we selected
line-free regions of the spectrum (mostly above 2--3 keV), rebinned
the data heavily, and fitted them with a smooth curve.  Second, we
used the zeroth-order spectrum from an extraction radius consistent
with the extraction width of the HETGS spectrum ($\sim$ 3\arcsec). The
latter was fitted with a power law plus free $N_{\rm H}$, plus narrow
(width=0.05 keV) Gaussians representing all the emission lines
detected in the HETGS spectrum. Both methods gave consistent
results. The 3-7 keV continuum can be described by an inverted power
law with photon index $\Gamma \sim -0.8$ and flux F$_{2-10~keV} \sim 7
\times 10^{-12}$ \flux.  This continuum was used in the EW
measurements.  The uncertainty on the continuum flux is 22\% or better
at all wavelengths.
  
\section{Discussion} 

We modeled the observed spectrum using our new line measurements, as
well as the information obtained from the radial flux distribution \
(from our companion paper) \ and the high en-  

\begin{table*}[t]
\footnotesize
\caption{X-ray Emission Lines
\label{observations}}
\begin{center}
\begin{tabular}{lccccc}
\hline
\hline
{Line and energy(eV)} &
{Wavelength} &
{Energy} &
{Flux} &
{EW} &
{EW} \\
{} &
{(\AA)} &
{(keV)} &
{($\times 10^{-4}$ ph cm$^{-2}$ s$^{-1}$)} &
{(eV)} &
{(\AA)} \\
{(1)} &
{(2)} &
{(3)} &
{(4)} &
{(5)} &
{(6)} \\
\hline
\ion{Ne}{10} (1211) & 10.236 & 1.211 $\pm$ 0.003 & 0.025 $\pm$ 0.015 & 24 $\pm$ 17 & 0.20 $\pm$ 0.14 \\   
\ion{Mg}{11} (1352) & 9.171 & 1.352 $\pm$ 0.003 & 0.058 $\pm$ 0.017 & 63 $\pm$ 24 & 0.43 $\pm$ 0.16 \\
\ion{Mg}{11} (1343)  &  9.245 & 1.341 $\pm$ 0.003 & $0.026 \pm 0.013$ & $28  \pm 17$ & $0.20 \pm 0.12$\\
\ion{Mg}{11} (1333)  &  9.300 & 1.332 $\pm$ 0.003 & $0.034 \pm 0.016$ & $36  \pm 19$  & $0.25 \pm 0.13$\\
\ion{Mg}{12} (1472)  &  8.419 & 1.471 $\pm$ 0.004 & $0.065 \pm 0.016$ & $77  \pm 23$  & $0.44 \pm 0.13$\\
\ion{Mg}{11} (1579)  &  7.851 & 1.578 $\pm$ 0.004 & $0.019 \pm 0.013$ & $23 \pm 13$  & $0.12 \pm 0.08$\\
\ion{Mg}{12} (1745)  &  7.106-7.130 & 1.741
$\pm$ 0.005 & $0.086 \pm 0.017$ & $116 \pm 32$  & $0.47 \pm 0.13$\\
+ \ion{Si}{2}--\ion{}{6}(1740--1746) &&&&& \\
\ion{Si}{13} (1865)  &  6.648 & 1.863 $\pm$ 0.006 & $0.087 \pm 0.016$ & $120
\pm 23$  & $0.43 \pm 0.08$\\ 
\ion{Si}{13} (1839)   & 
6.74 + 6.737 & 1.841 $\pm$ 0.006 & $0.105 \pm 0.018$ & $146 \pm 38$  & $0.54
\pm 0.14$\\ 
+ \ion{Mg}{12} (1840) &&&&& \\
 \ion{Si}{14} (2007)  &  6.180 & 2.006 $\pm$ 0.007 & $0.064 \pm
0.015$ & $91  \pm 28$  & $0.28 \pm 0.09$\\ 
\ion{S}{2}--\ion{}{10}(2308--2350)    &  5.28--5.37 & 2.313
$\pm$ 0.009 & $0.029 \pm 0.032$ & $43  \pm 47$ & $0.10 \pm 0.11$\\
 \ion{Si}{14} (2377)    &  5.217 + 5.194 & 2.383 $\pm$ 0.009 & $0.048
\pm 0.037$ & $71  \pm 46$  & $0.16 \pm 0.10$\\ 
+ \ion{S}{12} (2387) &&&&&  \\ 
 \ion{S}{14} (2411)   &  5.142
& 2.412 $\pm$ 0.009 & $0.028 \pm 0.027$ & $ 42 \pm 40$  & $0.09 \pm 0.08$\\ 
\ion{S}{15} (2430)   &  5.100 & 2.433 $\pm$ 0.010 & $0.056
\pm 0.029$ & $83  \pm 39$  & $0.18 \pm 0.09$\\ 
\ion{S}{15} (2461)   &  5.039 &
2.458 $\pm$ 0.010 & $0.085 \pm 0.036$ & $125 \pm 61$  & $0.26 \pm 0.13$\\
\ion{S}{16} (2623) & 4.727 & 2.624 $\pm$ 0.011 & 0.022 $\pm$ 0.023 & 33 $\pm$
35 & 0.06 $\pm$ 0.06 \\ 
\ion{S}{15} (2883) & 4.299 & 2.882 $\pm$ 0.011 & 0.021
$\pm$ 0.023 & 31 $\pm$ 34 & 0.05 $\pm$ 0.05 \\
 \ion{Ar}{2}--\ion{}{11} (2962--2990) & 4.12--4.187 & 2.960$\pm$ 0.014  & $0.06 \pm 0.031$ & $83 \pm 40 $ & $0.12 \pm 0.06$\\  
\ion{Ar}{17} (3104--3140) &  3.949--3.994 & 3.683
$\pm$ 0.022 & $0.039 \pm 0.021$ & $54.2  \pm 28.7$  & $0.05 \pm 0.03$\\
\ion{Ar}{18} (3323) & 3.794 & 3.268 $\pm$ 0.020 & 0.040 $\pm$ 0.024 & 59 $\pm$
38 & 0.07 $\pm$ 0.05 \\ 
 \ion{Ar}{18} (3936) &  3.150 & 3.917 $\pm$ 0.025 &
$0.041 \pm 0.026$ & $55  \pm 31$  & $0.05 \pm 0.03$\\ 
\ion{Ca}{2}--\ion{}{14}(3690--3760)  & 3.30--3.36 & 3.683 $\pm$ 0.022 &
0.038 $\pm$ 0.021 & 53 $\pm$ 31 & 0.05 $\pm$ 0.03 \\ 
+ \ion{Ar}{17} (3684) &&&&&  \\
 \ion{Ca}{20} ? (5115) &
2.424 & 5.166 $\pm$ 0.042 & 0.102 $\pm$ 0.040 & 127 $\pm$ 56 & 0.06 $\pm$ 0.03
\\ 
 ?          &  2.606 & 4.758 $\pm$ 0.037 & $0.052 \pm 0.029$ & $64 \pm 38$ 
& $0.04 \pm 0.02$\\  
?          &  2.069 & 5.993 $\pm$ 0.058 & $0.145 \pm
0.054$ & $177 \pm 69$  & $0.06 \pm 0.02$\\ 
\ion{Fe}{2}--\ion{}{17}(6400) &  1.942 &
6.385 $\pm$ 0.066 & $3.047 \pm 0.173$ & $2108 \pm 479$ & $0.64 \pm 0.15$\\
\ion{Fe}{25} (6637-6700)  &  1.851--1.868 & 6.656 $\pm$ 0.071 & $0.273 \pm
0.110$ & $211 \pm 91 $ & $0.06 \pm 0.03$\\ 
\ion{Fe}{2}--\ion{}{17}(7080--7196)   & 
1.72--1.75 & 7.030 $\pm$ 0.080 & $0.680 \pm 0.153$ & $602 \pm 189 $ & $0.15
\pm 0.05$\\ 
\hline
\end{tabular}
\end{center}
\end{table*}
\normalsize

\noindent ergy observations from \sax\
(Matt et al. 1999) and \rxte.  The modeling is based on the following
observations: (a) the 2--10 \AA\ continuum is flat, in $F_{\lambda}$,
at long wavelengths (as expected from an ``ionized mirror'') and
hardens below about 5\AA\ (as expected from a ``neutral
mirror''). This, plus the detection of low ionization species,
suggests two distinct components; (b) the EW of the \feka\ line is
very large, indicating iron over abundance (e.g. Netzer et al. 1998);
(c) the extended spectrum (outside of the central 0.8\arcsec) is
flatter than the central spectrum, suggesting that the more neutral
component contributes less at larger radii.

Modeling is done using ION00, the 2000 version of the photoionization
code ION (Netzer 1996).  This includes the computation of the steady
state ionization and thermal structure of the gas, and the emergent
spectrum. The main ingredients of the model are:
%
(a) Central power-law continuum with $\Gamma=1.5$, extending from
0.1--100 keV and normalized to produce $L_{2-10~{\rm keV}}=10^{42}$
\lum, in agreement with the \sax\ and \rxte\ observations;
%
(b) two component gas with twice-solar metallicity. We have
experimented with the density and column density of the two and found
the following satisfactory combination: 1. An {\it ionized component}
with a radial density distribution of $N(r)=N_0(r/r_0)^{-1.5}$ where
$r_0=1$ pc and $N_0=2 \times 10^3$ cm$^{-3}$. This distribution is
consistent with our measured radial flux distribution assuming most of
the 0.8--3\arcsec\ flux is due to scattered continuum.  2. A {\it
neutral component} with a similar (yet unconstrained) radial
distribution with the same $r_0$, but with $N_0=2 \times 10^5$
cm$^{-3}$. The first component is allowed to extend all the way to 300
pc while the second is limited by its column density, arbitrarily
chosen at $10^{23.9}$ cm$^{-2}$, and is thus terminated well inside
the inner 8 pc.  This component can perhaps be viewed as the wall of
the inner torus. The covering factors of the two components are free
parameters of the model;
%
(c) the gas turbulent velocity can range from no turbulence (pure
thermal motion) up to $few \times 100$ km s$^{-1}$. The increased line
width results in an increased intensity of all resonance lines due to
continuum fluorescence (Krolik and Kriss 1995; Netzer 1996). Our best
model requires no turbulent motion.

Several models have been computed, with various covering factors. A
satisfactory solution is found for $\Omega/4 \pi (neutral)$ =
0.4--0.5, and $\Omega/4 \pi$ (ionized) = 0.1--0.2, consistent with the
expected opening angle of the (hypothetical) torus, where $\Omega$ is
the solid angle subtended by the gas to the illuminating
source. Figure \ref{model-data} shows the two components alongside
with the observations and Figure \ref{ratio} shows observed over
computed intensities for the strongest lines.  Note that, given the
gas density and location, the only free parameters are the covering
factors of the two components.

The overall agreement between the model and the observations is good,
given the uncertainties. In particular: (a) the observed fluxes of
most emission lines are reproduced, to within a factor of two. The
1.5--11\AA\ continuum shape is reproduced too. These \ results \ support
the \ idea of the \ two component \ model,

\centerline{\epsfxsize=10.5cm\epsfbox{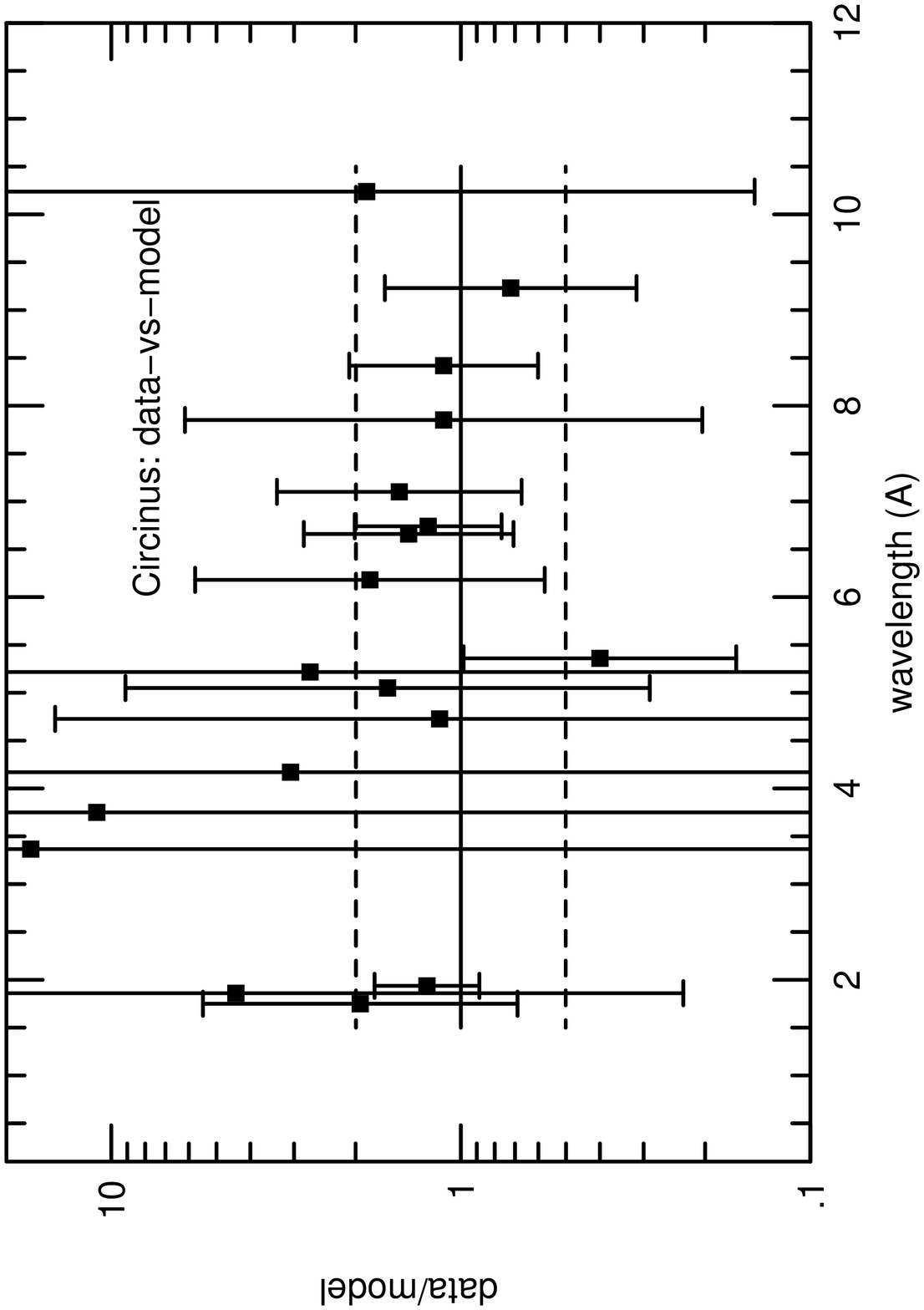}}
\figcaption{
A comparison of observed and calculated line intensities. The
lines can be identified by their wavelengths. The solid line marks the
locus of equality between data and model, the dashed lines represent a 
factor 2 uncertainty from the solid line. 
\label{ratio} }
\centerline{}
\centerline{}

\noindent with the assumed levels of
ionization and metallicity; (b) the covering fractions are consistent
with the suggested source geometry. The radial density distribution
reproduces well the highly peaked emission of this source and the
relative weak flux outside the central 16 pc. The calculated scattered
continuum and extended emission lines in the inner 16--57 pc are in
good agreement with the (highly uncertain, see our companion paper)
observations and there is no need to assume an additional starburst
source; (c) the ``neutral'' iron K$\alpha$ and K$\beta$ lines are
consistent with the compact, neutral component gas and the assumed
metallicity. Notable difficulties are the over-prediction of the Fe
L-shell lines around 9--12 \AA\ and the under-prediction of the 6.6
keV feature if due to \ion{Fe}{25}. All argon lines are also
under-predicted by the model.  The measured intensity of these lines
are highly uncertain and any suggestion for their origin (e.g. unusual
composition) must await better observations.  We also note that the
predicted K$\alpha$ flux, outside the central 20 pc, is below the
observed value. This may be due to the already noted uncertainty in
flux measurement.  We also note that a ``typical'' NLR, can produce a
sizeable fraction of this K$\alpha$ emission.

Regarding earlier X-ray observations of this source, the discovery
paper by Matt et al. (1996) reports an \asca\ spectrum including both
neutral and ionized species.  Netzer et al. (1998) re-analyzed the
\asca\ data and reported the measurements of six ionized lines plus
the iron K-lines, all in reasonable agreement with the present
observations.  The paper includes a detailed photoionization model of
the source and addresses also the K$\alpha$/H$\beta$ line ratio.
Guainazzi et al. (1999) reported on \sax\ data that include the
measurement of \ion{Si}{13}, \ion{S}{15}, and \ion{Ar}{17} lines, as
well as neutral Fe-K lines. The observed fluxes are in good agreement
except for \ion{Ar}{17} whose \sax\ intensity exceeds our estimate by
a factor 4. 

The present observations, being far superior in terms of the spatial
and spectral resolution, yet limited in signal-to-noise ratio, confirm
several of the suspected features of this source, such as the gas
location, metallicity, and the various components.  This, and the
recently published \chandra\ observations of Mkn~3 (Sako et al. 2000b)
show, for the first time, that centrally illuminated ionized gas in
Seyfert 2 galaxies can have very different distributions in different
sources. While in Mkn~3 most of the line emission is spread over
several hundred parsecs, this in not the case in \cir, where the
ionized gas is highly concentrated near the center and most of its
flux originates within the central $\sim 15$ pc. Moreover, the new
observations show that the more neutral gas is even more
concentrated, and its dimension may be as small as a few parsecs.
While we do not want to speculate about the origin of this difference,
we note the large difference in luminosity between these two
sources. It is therefore possible that the highly ionized gas in
Seyfert galaxies has dimensions that are regulated by the central
source's X-ray luminosity. 

\section{Conclusions}

The new \chandra\ observations of the Seyfert 2 galaxy \cir\ enable
the very first detailed analysis of the physical conditions and the
gas distribution in the inner 200 pc of this source.  Our observations
show the emission to be highly concentrated within the inner 60 pc and
suggest that emission within this volume is entirely due to the
reprocessing of the obscured central source's radiation.  An even
smaller, more neutral component is seen through emission of low
ionization iron lines and hard reflected continuum.  This is the first
determination of the ionized gas distribution in the inner 100 pc
region of a Seyfert 2 galaxy.

\acknowledgements 

We acknowledge the financial support of NASA grant NAS8--38252 (RMS; GPG
PI), NASA grant NAG5--7276 from \rxte\ AO3, NASA LTSA grant NAG5--8107
(WNB, SK), and  the Israel Science Foundation (HN). 


\end{document}